\documentclass{elsarticle}

\usepackage{color}
\usepackage{lineno}

\begin{document}


\begin{frontmatter}

\title{A liquid helium target system for a measurement of parity violation in neutron spin rotation}

\author[IU,NIST]{C. D. Bass\corref{cor1}}
\ead{christopher.bass@nist.gov}
\author[IU]{T. D. Bass}
\author[UW]{B. R. Heckel}
\author[IU]{C. R. Huffer}
\author[IU]{D. Luo}
\author[NCCU]{D. M. Markoff}
\author[IU]{A. M. Micherdzinska}
\author[IU]{W. M. Snow}
\author[UW]{H. E. Swanson}
\author[IU]{S. C. Walbridge}
\address[IU]{Indiana University / IUCF, Bloomington, IN 47408, USA}
\address[NIST]{National Institute of Standards and Technology, Gaithersburg, MD 20899, USA}
\address[UW]{University of Washington / CENPA, Seattle, WA 98195, USA}
\address[NCCU]{North Carolina Central University, Durham, NC 27707, USA}
\cortext[cor1]{Present address: National Institute of Standards and Technology, Gaithersburg, MD 20899.}

\begin{abstract}
A liquid helium target system was designed and built to perform a precision measurement of the parity-violating neutron spin rotation in helium due to the nucleon-nucleon weak interaction.  The measurement employed a beam of low energy neutrons that passed through a crossed neutron polarizer--analyzer pair with the liquid helium target system located between them.  Changes between the target states generated differences in the beam transmission through the polarizer--analyzer pair.  The amount of parity-violating spin rotation was determined from the measured beam transmission asymmetries.  The expected parity-violating spin rotation of order $10^{-6}$ rad placed severe constraints on the target design.  In particular, isolation of the parity-odd component of the spin rotation from a much larger background rotation caused by magnetic fields required that a nonmagnetic cryostat and target system be supported inside the magnetic shielding, while allowing nonmagnetic motion of liquid helium between separated target chambers.  This paper provides a detailed description of the design, function, and performance of the liquid helium target system.
\end{abstract}


\end{frontmatter}

\section{Introduction}
The Neutron Spin Rotation experiment was conducted to measure the parity-violating spin rotation angle $\phi_{\rm{PV}}$ of cold neutrons that propagate through liquid helium to a precision of $3 \times 10^{-7}$ rad/m.  The liquid helium target system for this experiment was assembled and tested at the Indiana University Cyclotron Facility (IUCF) in Bloomington, Indiana, and the experiment was conducted at the National Institute of Standards and Technology (NIST) in Gaithersburg, Maryland.

This paper describes the design, commissioning and performance of the liquid helium target system used in the Neutron Spin Rotation experiment.  A more detailed description of the polarimeter and neutron beam characterization for this experiment will be addressed in future papers.  Section 1 provides a description of the phenomenon of parity violation in neutron spin rotation with a brief discussion of the scientific interest in its measurement; explains the design of the neutron polarimeter and the measurement strategy for isolating the parity-odd component of the neutron spin rotation; and describes the overall experimental apparatus, listing the requirements and constraints on the design of the liquid helium target. Section 2 specifies the design and construction of the liquid helium target system.  Section 3 outlines the motion control system that was used to move liquid within the target region.  Section 4 describes the nonmagnetic cryostat and the cryogenic performance of the target system. Section 5 specifies the integration of the liquid helium target system into the data acquisition system.  Section 6 provides details of the performance of the target, and Section 7 offers conclusions.

\subsection{Physics Overview}

From an optical viewpoint \cite{Mic64}, neutron spin rotation is caused by the presence of a helicity-dependent neutron index of refraction $n$ of a medium, which can be given in terms of the coherent forward scattering amplitude $f(0)$ for a low-energy neutron:

\begin{equation}
n=1-\frac{2\pi \rho f(0)}{k^{2}},
\end{equation}

\noindent where $\rho$ is the number density of scatterers in the medium, and $\vec{k}$ is the incident neutron wave vector.  For low-energy (meV) neutrons propagating through an unpolarized medium, $f(0)$ is the sum of an isotropic, parity-even term $f_{\rm{PC}}$ that is dominated by the strong interaction and a parity-odd term $f_{\rm{PV}}$ that contains only weak interactions and is dominated by p-wave contributions.  The $f_{\rm{PV}}$ term is proportional to $\vec{\sigma_{\rm{n}}}\cdot \vec{k}$, where $\vec{\sigma_{\rm{n}}}$ is the neutron spin vector, so $f_{\rm{PV}}$ has opposite signs for the positive and negative helicity neutron spin states.

As a neutron propagates through a medium, the two helicity states accumulate different phases: $\phi_{\pm}=\phi_{\rm{PC}} \pm \phi_{\rm{PV}}$.  The parity-odd component causes a relative phase shift of the two neutron helicity components, and so induces a rotation of the neutron polarization about its momentum.  Because the parity-odd amplitude is proportional to the wave vector $k$, the rotary power (rotation per unit length) tends to a constant for low energy neutrons \cite{Sto74}:

\begin{equation}
\lim_{E_{n}\to0}\frac{d\phi_{PV}}{dz}=\frac{4\pi\rho f_{PV}}{k}.
\end{equation}

\noindent An order-of-magnitude estimate leads one to expect weak rotary powers in the $10^{-6}-10^{-7}$ rad/m range.

The parity-violating neutron spin rotation is due to the nucleon-nucleon (NN) weak interaction and can be described in terms of the Desplanques, Donoghue, and Holstein nucleon-meson weak coupling amplitudes \cite{DDH80}\cite{Hol05} as well as pionless effective field theory coupling parameters \cite{Ram06}\cite{Liu05}\cite{Zhu05}.  The values of these couplings are neither well-constrained by theory nor by experiment, so a measurement of the parity-violating neutron spin rotation through liquid helium can constrain the poorly-understood properties of the NN weak interaction.

\subsection{Measurement Technique}

\begin{figure*}
    \centering
    \includegraphics[width=\linewidth]{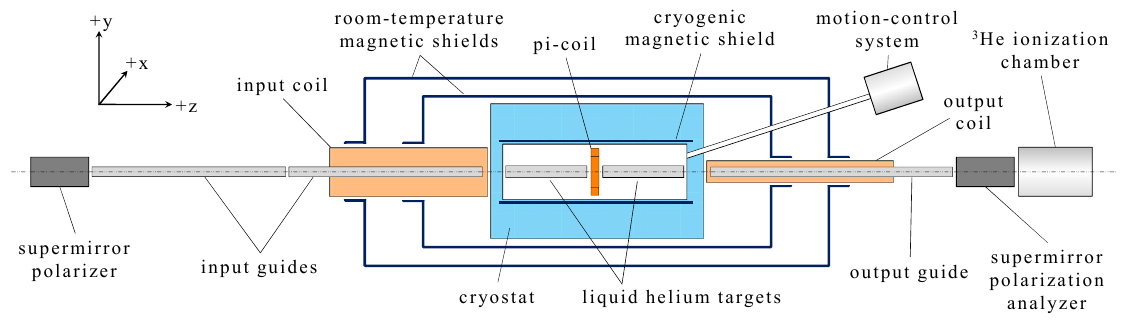}
    \caption{Schematic diagram of the neutron polarimeter apparatus for the Neutron Spin Rotation experiment.}
    \label{polarimeter}
\end{figure*}

An overview of the neutron polarimeter is shown in Figure~\ref{polarimeter}.  The measurement technique -- analogous to an arrangement of an orthogonally-crossed polarizer--analyzer pair in light optics -- focuses on the orientation of the neutron polarization, which emerges along the $+\hat{\rm{y}}$-direction from the supermirror polarizer \cite{Hec81}.

\begin{figure*}
    \centering
    \includegraphics[width=\linewidth]{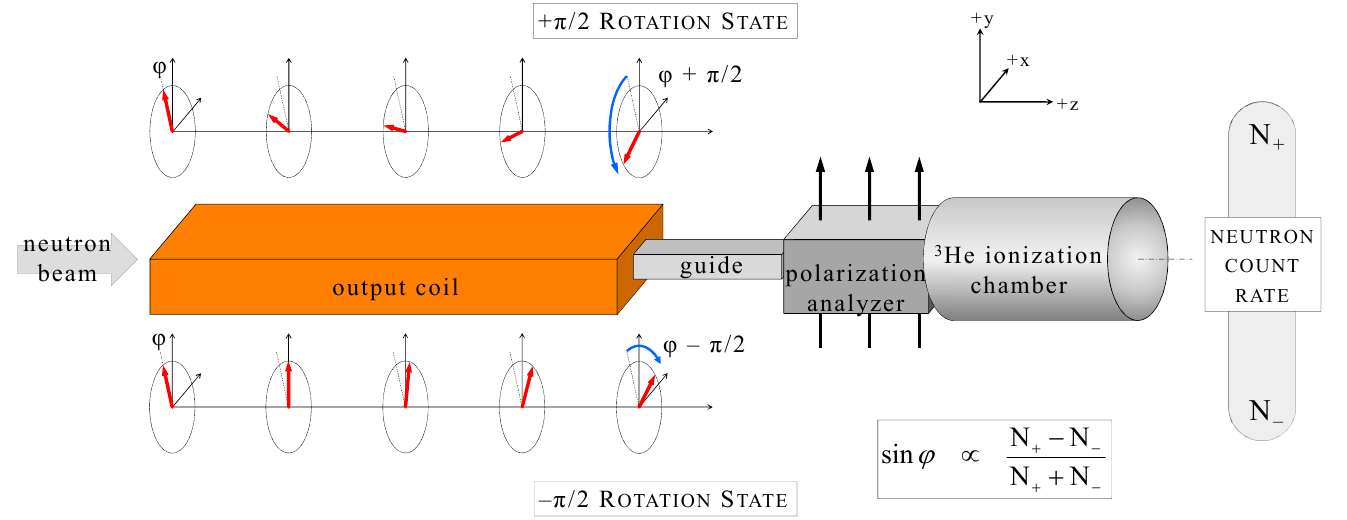}
    \caption{Diagram of the transverse rotation of neutron polarization by the output coil. The amount of spin rotation experienced by the neutrons along the target region is proportional to the count rate asymmetry measured in the $^{3}$He ionization chamber between the two output coil rotational states.}
    \label{output coil}
\end{figure*}

In the absence of spin rotation -- including both background magnetic rotations and the signal rotations -- this orientation would remain unchanged during passage along the spin transport and into the target region.  After leaving the target region, neutrons would enter the output coil, which transversely and adiabatically rotates the neutron polarization vector by $\pm \pi / 2$ rad (see Figure~\ref{output coil}). Neutrons would then pass through the polarization analyzer.  Because the transmitted beam intensity for both $+ \pi / 2$ and $- \pi / 2$ rotational states is the same, the difference in count rates measured by the $^{3}$He ionization chamber would be zero.

However, if the neutron polarization rotates during beam passage through the target region, there would be a component of neutron polarization along the $\hat{\rm{x}}$-direction (horizontal) when the beam reaches the output coil. This component would flip between the $+\hat{\rm{y}}$ and $-\hat{\rm{y}}$-directions as the output coil alternates between $+\pi / 2$ and $-\pi / 2$ rotational states.  The transmission of neutrons polarized parallel to the axis of the polarization analyzer would be different than those polarized antiparallel, and this would produce an asymmetry in the count rates for the two output coil rotational states. The neutron spin-rotation angle $\phi_{\rm{PV}}$ is proportional to this count rate asymmetry.

Because longitudinal magnetic fields generate neutron spin rotation, it was necessary to separate the parity-violating component of the signal from this parity-conserving background.  This separation was accomplished by oscillating the parity-violating signal at a known frequency.

However, since the magnitude and direction of spin rotation along the target is independent of the initial direction of the neutron transverse polarization, flipping the direction of the transverse polarization cannot be used to modulate the parity-violating signal.  Instead, the oscillation was created by a combination of target motion and a precession of the neutron polarization about a vertical axis.  Liquid helium was moved between a pair of target chambers located upstream and downstream of a vertical solenoid called a ``pi-coil''.

The integrated spin rotation in the target region due to magnetic fields was unaffected by the presence of liquid helium in either the upstream or downstream target chambers, provided that (1) the target was nonmagnetic and (2) the trajectories and energies of neutrons accepted by the polarization analyzer and $^{3}$He ionization chamber were unchanged when liquid helium was moved between the target chambers.

\begin{figure*}
    \centering
    \includegraphics[width=\linewidth]{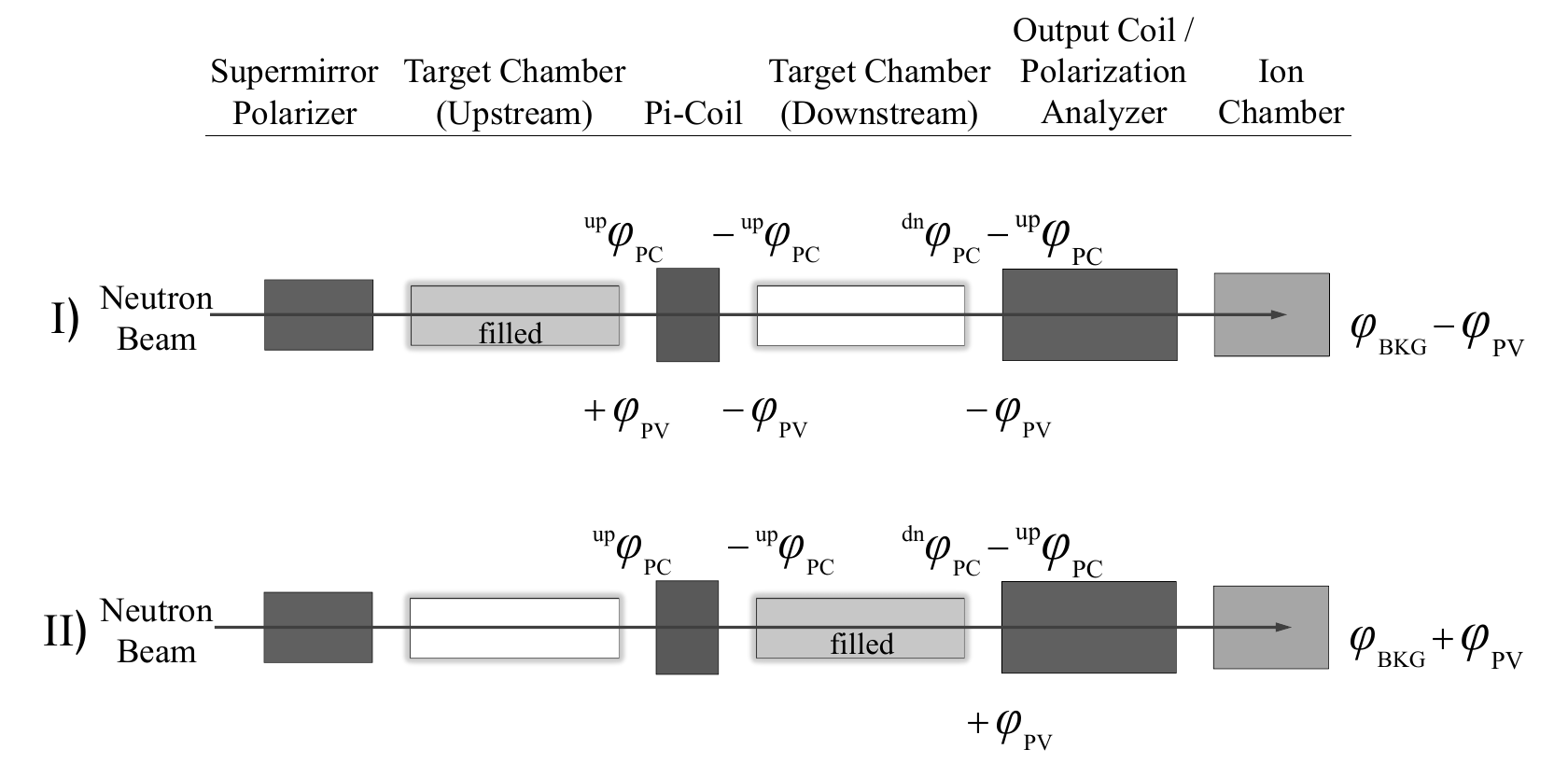}
    \caption{Diagram of the spin rotation angles along the polarimeter.  As neutrons pass through the upstream section of the target, they undergo a parity-conserving Larmor precession $^{up}\phi_{PC}$, which is due to background magnetic fields in that region.  Passage though the pi-coil reverses the sign of that rotation.  Neutrons then pass through the downstream target region and undergo another parity-conserving Larmor precession $^{dn}\phi_{PC}$ of similar magnitude as $^{up}\phi_{PC}$ due to fields in that region.  The residual background rotation $\phi_{bkg}$ is the small difference between $^{up}\phi_{PC}$ and $^{dn}\phi_{PC}$, and is independent of the presence of helium in the target chambers.  However, neutrons passing through the target when the upstream chamber is filled with liquid helium (I) undergo an additional parity-violating spin rotation $\phi_{PV}$ that is due to the NN weak interaction.  The sign of $\phi_{PV}$ is reversed upon passage through the pi-coil.  Similarly, neutrons passing through the target when the downstream chamber is filled with liquid helium (II) also undergo a parity-violating spin rotation $\phi_{PV}$.  However, these neutrons have already passed through the pi-coil, so the sign of $\phi_{PV}$ is unchanged.  Thus, moving liquid helium between upstream and downstream target chambers can modulate the sign of $\phi_{PV}$ signal against a constant background $\phi_{bkg}$.}
    \label{schematic}
\end{figure*}

The pi-coil generated an internal magnetic field whose magnitude was chosen to precess the neutron polarization direction by $\pi$ radians about the $\hat{\rm{y}}$-axis for neutrons of a given energy.  This precession effectively reversed the sign of the x-component of neutron spin for rotations that occurred upstream of the pi-coil. With the pi-coil energized, the parity-violating contribution to the neutron spin rotation for liquid helium was negative for the target chamber located in the upstream position relative to the rotation in the downstream target position, and the difference in the total spin rotation angle between the two target states was $2\phi_{\rm{PV}}$ plus a contribution due to non-ideal magnetic backgrounds within the target region. A schematic of the spin rotation angles along the polarimeter is shown in Figure~\ref{schematic}.

\subsection{Physics Driven Parameters}
In order to reach a measurement sensitivity of $3 \times 10^{-7}$ rad/m, the neutron polarimeter and liquid helium target system needed to satisfy certain requirements.  Because the parity-violating rotary power is constant in the limit for low-energy neutrons (Equation 2), the parity-odd spin-rotation angle $\phi_{PV}$ is proportional to the number density of scatterers $\rho$ for the target and the target length $z$.  The statistical error is given by

\begin{equation}
\sqrt{N} = \sqrt{N_{o}e^{-\rho\sigma_{tot}z}},
\end{equation}

\noindent with total neutron cross section $\sigma_{tot}$.  Maximizing the signal to statistical error yields a target length of $z=2/\rho\sigma_{tot}$ or twice the mean free path.  Because $\sigma_{tot}$ strongly depends on energy for liquid helium in the cold neutron regime, the optimal target length depends on the details of the neutron energy spectrum.  In liquid helium at 4.2 K, the mean free path for 0.5 nm neutrons is about 1 meter, and this path length increases to about 2 m for 0.7 nm neutrons \cite{So55}.

The polarimeter needed to effectively isolate the parity-violating spin rotation signal from noise and a much larger parity-conserving background.  The primary sources of noise above $\sqrt{N}$ counting statistics were (1) fluctuations in the neutron beam intensity that were transmitted through the polarization analyzer and into the ion chamber, which were due to noise in the intensity and/or beam spectrum from the reactor, or from transmission through density fluctuations in the liquid helium target; (2) fluctuations of the neutron spin rotation angle that were analyzed by the polarimeter, which were caused by magnetic field fluctuations in the target region; and (3) extra noise in the ion chamber due to the current-mode measurement technique.  Noise from (1) was relevant to the target design and is discussed, while the others sources will be covered in a later paper on the apparatus as a whole.

The two relevant frequency bands for the Neutron Spin Rotation experiment are the modulation frequency of the output coil (which determined the measurement frequency of the neutron spin rotation angle from count-rate asymmetry measurements) and the frequency of the target motion (which determined the measurement frequency of the parity-violating component of the spin rotation angle). In general, these frequencies should be as high as possible. The modulation frequency of 1 Hz for the output coil was chosen by how quickly the currents in the coil could be reversed and stabilized.  The target motion frequency of about 1 mHz was set by the performance of the liquid helium pump and drain system, and the volume of liquid helium to be moved.

Although no precise measurements of reactor intensity or spectrum fluctuations had been performed for the NIST NG-6 cold neutron beam in the 1 Hz frequency range before this measurement, a time-series analysis of neutron flux monitor data from a neutron lifetime experiment (sampled at $60$ second intervals) \cite{Ni06} combined with an assumed $1/f$ dependence of the source noise that has been observed in other reactors, implied that the beam intensity noise in a 1 Hz frequency band could be $5-10$ times larger than the $\sqrt{N}$ noise from the integrated number of neutrons in the same band.  This reactor noise could be suppressed by segmenting both of the target chambers into separate left and right halves and then splitting the neutron beam into two parallel sub-beams, which effectively created two separate simultaneous experiments.  Operation of the left and right side targets with opposite target states allowed a comparison of left and right side asymmetry measurements that suppressed common mode noise.

Constraining the common-mode noise associated effects to be smaller than neutron counting statistics required that the two parallel sub-beams possess the same intensity noise to 1\% accuracy.  Since the phase space of both halves of the beam was filled from the same neutron source as viewed through a long, uniform neutron guide, the common-mode intensity fluctuations for each half of the beam were expected to possess similar power spectra.

The energy spectrum for neutrons entering the polarization analyzer was primarily determined by the cold neutron moderator, the phase space acceptance of the guides, and the beam transmission through the liquid target.  None of these elements was likely to shift the energy spectrum by more than 0.1\% at 1 Hz, so this contribution to the noise was expected to be negligible.  Fluctuations in the attenuation of the beam through the target also increased the noise, but stability of the liquid helium density at the 1\% level in the frequency band of interest near 1 Hz was not difficult to achieve.

Finally, since most sources of systematic uncertainties scaled with the strength of the longitudinal magnetic field in the target region, suppression of these effects to below the $5 \times 10^{-8}$ rad/m level required a longitudinal magnetic field of less than 10 nT in the target region.

\subsection{Overall Design Considerations and Constraints}

The neutron polarimeter employed a target system that moved liquid helium between a pair of target chambers located upstream and downstream of a vertical solenoid in order to generate beam count-rate asymmetries, which were used to determine spin rotation angles.  In the ideal case of no target-dependent magnetic rotations, the location of liquid helium in the target should generate the parity-violating signal and not affect the parity-conserving background signal.  The transfer of liquid helium between the target chambers should change only the sign of the parity-violating signal but not its magnitude and must be carried out non-magnetically.

The presence of liquid helium in either the upstream or downstream target chamber defined a ``target state'' of the polarimeter.  The length of time that the liquid helium target remained in a given target state was determined by beam count-rate statistics and dead-time during target state change-overs.  These time parameters were related to the data acquisition dead-time and to possible systematic effects caused by time-dependent magnetic fields in the target region \cite{Ba108}.

In order to reduce the cost of the polarimeter and liquid helium target system, the collaboration chose to reuse several components from a previous experimental apparatus used to search for spin rotation in helium \cite{Mar97}.  These components included a nonmagnetic, horizontal bore cryostat; a pair of coaxial room-temperature magnetic shields with endcaps; input and output guide coils for the polarimeter; the pi-coil; and the liquid helium centrifugal pump.

The 100 cm long bore of the cryostat bound the overall length of the liquid helium target assembly, which included the targets and pi-coil, mechanical supports, a cylindrical vacuum vessel that housed the target assembly, and necessary vacuum hardware and electrical instrumentation feedthroughs.  In order to optimize available space, target chamber lengths were set to 42 cm.

The transverse dimensions of the target were determined by the cross-sectional area of the neutron beam at the exit window of the supermirror polarizer, which was 4.5 cm tall by 5.5 cm wide. The dimensions inside the target chambers allowed maximum acceptance of the beam with sufficient space for internal collimation. Combined with the target length, these dimensions determined the target volume.

The upstream and downstream target chambers were split vertically by a 3 mm thick septum, which created left and right-side chambers.  The collimation within the target chambers spit the neutron beam into separate but parallel sub-beams, which effectively created two separate simultaneous experiments.  The left and right-side targets were operated with opposite target states, so that the parity-violating signal was opposite in sign for the left and right sides while the parity-conserving background for both sides was the same.  Comparison of left and right-side measurements allowed a suppression of systematic effects and common-mode noise as described in Section 1.2.

The need to suppress the magnetic field inside the target region severely affected the design of the liquid helium target system and the spin transport of the polarimeter.  Within the target region, only non-magnetic and low-permeable materials were used -- all hardware was checked explicitly for magnetic inclusions or impurities.  Any item that produced changes in the ambient magnetic field of more than 1 nT when moved past a fluxgate magnetometer sensor at a distance of 1 cm was rejected.

The current-carrying wires of the pi-coil and the instrumentation potentially generated undesired magnetic fields, therefore twisted-pair wires were used for all wiring within the target region in order to suppress associated stray fields.  Furthermore, all instrumentation was powered off during data-taking and energized during target changes as required.

\subsection{Experimental Layout}

Low energy ($\sim$10$^{-3}$ eV) neutrons from the NIST Center for Neutron Research (NCNR) cold source were transported to the end station of the NG-6 polychromatic beam line and passed through cryogenic blocks of bismuth to filter out gamma rays and fast neutrons and beryllium to filter short wavelength cold-neutrons.  Transmitted neutrons passed into the neutron polarimeter apparatus as shown in Figure~\ref{polarimeter}.

At the upstream end of the apparatus, neutrons were vertically polarized in the ``up'' direction ($+\hat{\rm{y}}$-axis) by a supermirror polarizer.  The neutrons traveled along a 1.25 m long guide tube that was filled with helium gas and passed through an input coil, whose vertical field preserved the alignment of the beam polarization.  Neutrons passed through a current sheet at the end of the input coil so that the neutron spin was non-adiabatically transported into the magnetically shielded target region.

Reducing the ambient 50 mT field to less than 10 nT in the target region was accomplished using a combination of room-temperature and cryogenic mu-metal shielding.  Transient fields in the NCNR guide hall generated by equipment and other experiments were suppressed by compensation coils that were mounted outside the room-temperature mu-metal shielding.  Magnetometery located in the target region provided control feedback to the compensation coils.

After passing into the target region, neutrons were collimated into separate left and right sub-beams and allowed to enter the liquid helium target.  Neutrons propagated through the target and then entered the output coil, which non-adiabatically guided the neutron spin out of the target region.  The output coil also transversely rotated the neutron polarization vector by $\pm \pi / 2$ rad.  The sub-beams then passed through a supermirror polarization analyzer whose polarization axis was aligned with that of the supermirror polarizer.

Transmitted neutrons then entered a longitudinally-segmented $^{3}$He ionization chamber based on the design by Penn et al \cite{Pen01}, which operated in current mode and generated a signal proportional to the neutron count rate. The asymmetry in the count rates measured after changing the helicity of the magnetic transport field in the output coil was proportional to the neutron spin rotation angle.  The segmentation of the $^{3}$He ionization chamber provided some energy discrimination of the neutron beam.  The $1/v$ dependence of the neutron absorption cross-section in $^{3}$He allows higher energy neutrons to penetrate deeper on average into the ion chamber.  The parity-conserving spin rotation due to magnetic fields is proportional to the time neutrons spend in the fields and thus is dependent on neutron velocity.  The parity-conserving background should generate non-uniform asymmetries across the ion chamber segmentation that scale with the strength of the integrated residual magnetic field.  However, the parity-violating signal is independent of energy for cold neutrons and should generate the same asymmetry for all ion chamber segments.

\section{Liquid Helium Target}

The liquid helium target consisted of a pair of vessels located upstream and downstream of the pi-coil.  Each vessel was partitioned into separate left and right-side chamber pairs, which created four identical target chambers that could each hold liquid helium.  This partitioning effectively created upstream and downstream chamber pairs along the left and right-side sub-beams and formed two parallel experiments when combined with the polarimeter design such that whenever the upstream target chamber is filled, the corresponding downstream chamber is nominally empty and vice versa.  Alternately filling and draining diagonal pairs of target chambers with liquid helium created the two target states that were necessary to extract the parity-violating spin rotation angle $\phi_{\rm{PV}}$.

Each target chamber possessed an inlet and outlet for transferring liquid helium.  Inlets from all four chambers were connected to a centrifugal pump that was immersed in a 13 L liquid helium bath located in the bottom of a cylindrical vessel called the vac-canister.  By operating the centrifugal pump, all four target chambers could be filled with liquid helium.

\begin{figure*}
    \centering
    \includegraphics[width=\linewidth]{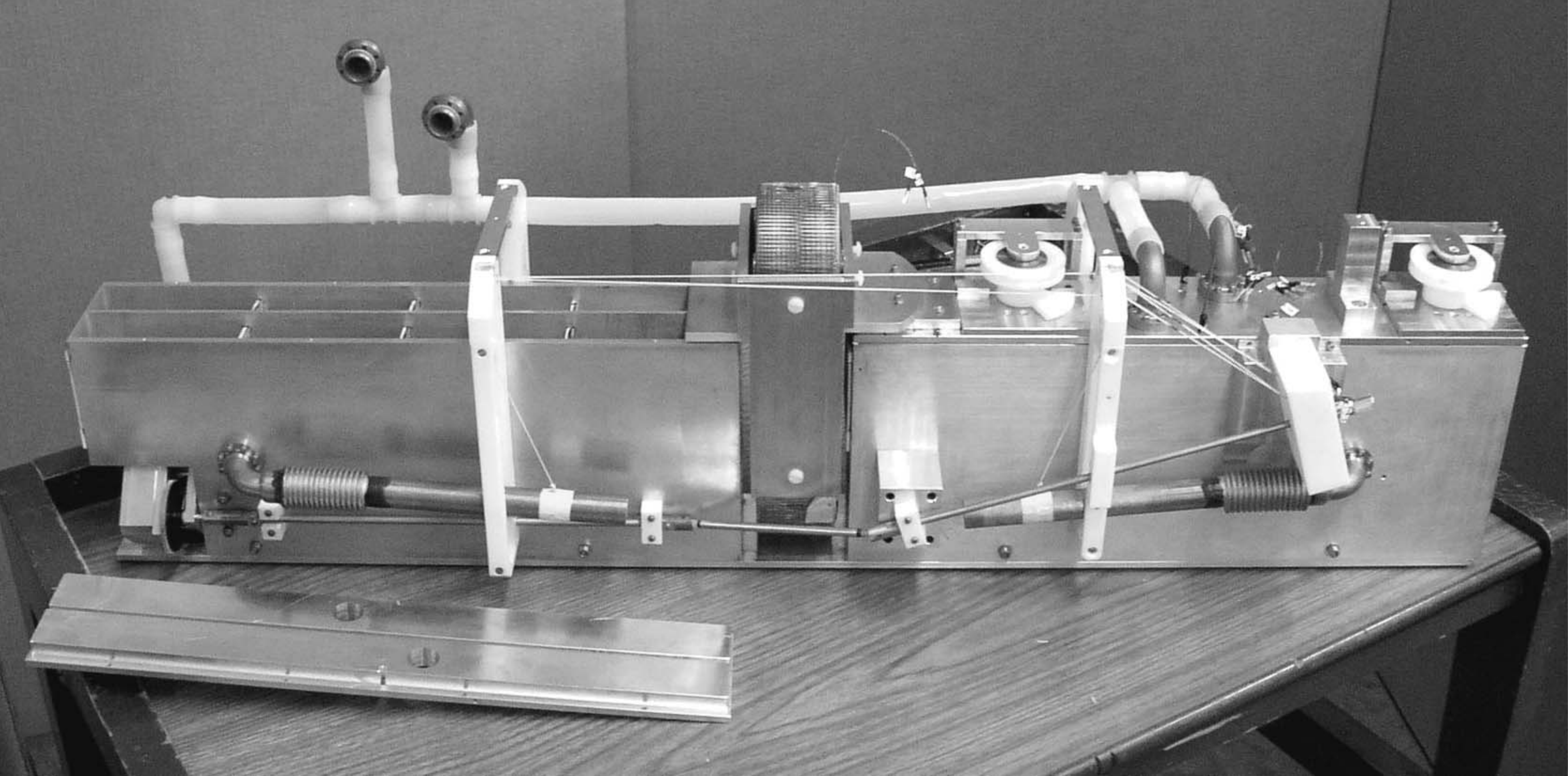}
    \caption{Photo of the liquid helium target showing the pump and drain system, the pi-coil, and instrumentation.}
    \label{chambers}
\end{figure*}

Each outlet was connected to a flexible drainpipe that could be moved above or below the height of nominally full or empty (respectively) liquid helium levels inside the target chambers.  By lowering a drain, a target chamber could be emptied of liquid helium and its contents returned to the bath.  The volume of the bath in the bottom of the vac-canister was maintained by periodic transfer of liquid helium from an external dewar.

\subsection{Vac-canister}

The vac-canister was a 95.3 cm long by 28.8 cm diameter cylindrical aluminum vacuum chamber that housed the liquid helium target.  Upstream and downstream targets -- as well as the pi-coil -- were bolted to an aluminum support rail that mated into a set of matching rails in the bottom of the vac-canister. This rail system provided alignment of the target within the vac-canister.

Main flanges with indium o-ring seals were located on the upstream and downstream ends of the vac-canister.  Both of the main flanges had a 0.8 mm thick aluminum flange with an indium o-ring seal covering a beam window.  During leak checks and data taking operation, the vac-canister and beam windows were shown to withstand an internal pressure difference of 270 kPa and an external pressure difference of 101 kPa without leak or rupture of the windows at both room temperature and cryogenic temperatures.  In addition, the downstream main flange was built to accept two custom-built nonmagnetic electrical feedthroughs \cite{Ba208}, a motion-control feedthrough tube, and a liquid helium transfer tube, along with spare access ports.  All seals used aluminum ConFlat\footnote{Certain commercial equipment, instruments, or materials are identified in this paper in order to specify the experimental procedure adequately. Such identification is not intended to imply recommendation or endorsement by the National Institute of Standards and Technology, nor is it intended to imply that the materials or equipment identified are necessarily the best available for the purpose.}(CF) flanges and gaskets or indium joints, and all metals were brass, aluminum or titanium.  All components were thermally cycled several times in preliminary tests and later were shown to be superfluid leak tight in low temperature tests at IUCF.

Around the outside of the vac-canister, ten equally spaced coils connected to individual current supplies provided magnetic field compensation inside a layer of cryogenic magnetic shielding that lined the cold bore.

\begin{figure}
    \centering
    \includegraphics[width=\linewidth]{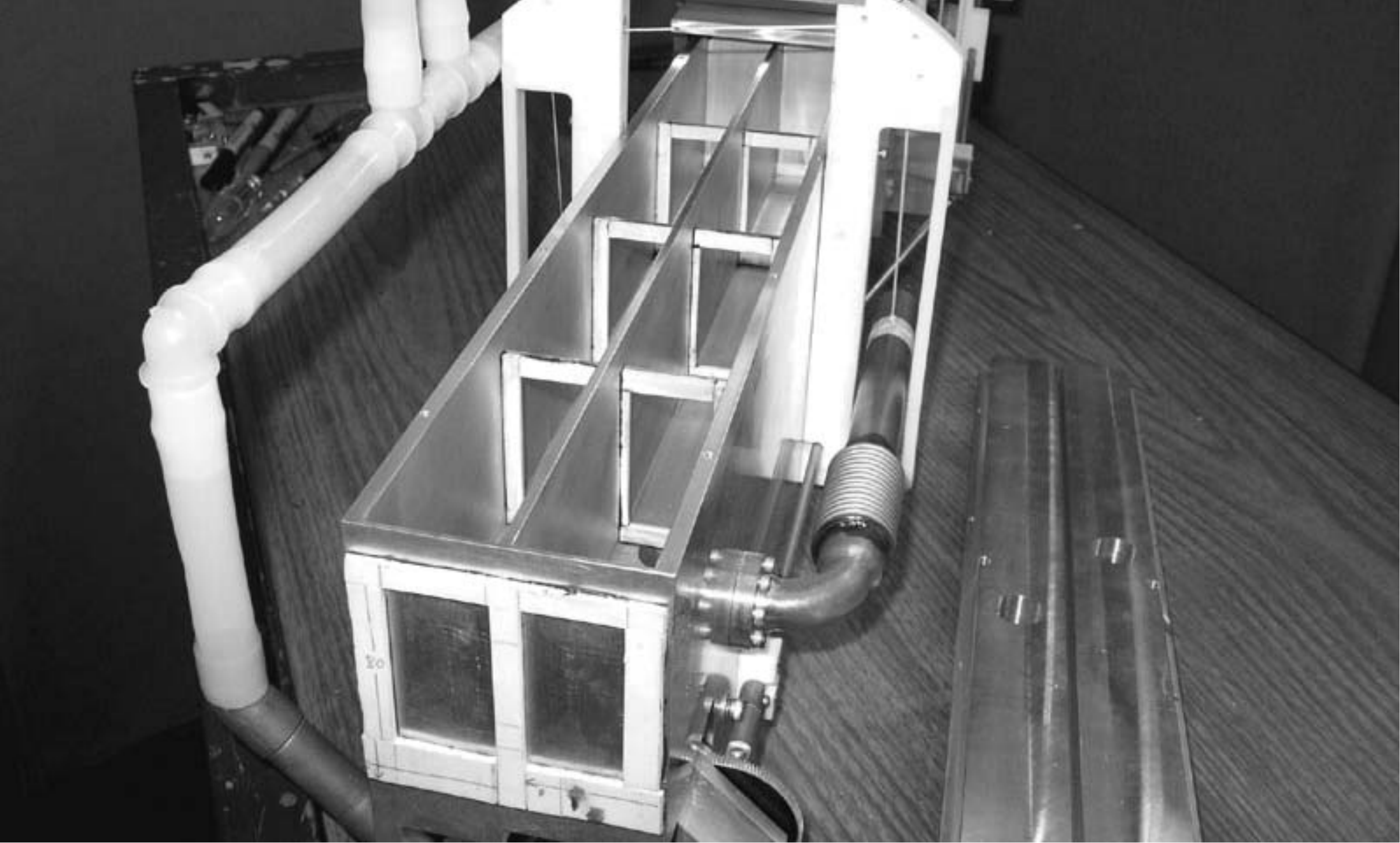}
    \caption{Layout of the neutron beam collimation within a target.  $^{6}$LiF-plastic was glued onto aluminum backing frames and then glued into the target chambers.  The size and spacing of the collimation was chosen to prevent neutrons with wavelengths larger than 2 nm from reflecting off a chamber wall and then being accepted into the $^{3}$He ionization chamber.  Additional $^{6}$LiF-plastic was glued onto the upstream face of the targets.}
    \label{collimators}
\end{figure}

The vac-canister rested within this cryogenic magnetic shielding on four ceramic balls, which thermally isolated it from the rest of the cryostat.  The vac-canister was able to slide along the cold bore to compensate for differential thermal contraction of the motion-control feedthrough and liquid helium transfer tubes, which connected the vac-canister to fixed connections on the motion-control box located outside the target region. A pair of titanium bolts protruded through the cryostat upstream 4-K thermal shield, which provided a low thermal conductivity mechanical stop for the upstream movement of vac-canister.

\subsection{Target Chambers}

The 420 mm long by 80 mm wide upstream and downstream targets were each machined from monolithic pieces of 6061 aluminum.  Aluminum was chosen as the target chamber material since it is nonmagnetic, has a high thermal conductivity, and has a low neutron scattering and absorption cross section \cite{Sea92}.

A wire electrical discharge machine (EDM) at the University of Washington was used to create left and right-side chambers in each target that were 416 mm long by 33.5 mm wide by 60 mm deep; the chambers were separated by a 3 mm thick septum that isolated the left and right sides (see Figure~\ref{chambers}).  Special care was taken to ensure that all surfaces exposed to the neutron beam were flat and normal to the mean beam direction to minimize possible systematic effects from neutron refraction.  In addition, special care was taken to make the target dimensions, especially the target lengths, as identical as possible to minimize target-dependent systematic effects.  The measured length difference at room temperature between all four target chambers was less than 0.01 mm.

\subsection{Neutron Beam Collimation}

Because of beam divergence or small angle scattering, some neutrons could reflect from a target chamber wall, be transmitted through the polarimeter, and counted in the $^{3}$He ionization chamber. The critical angle for neutron reflection between helium and aluminum depends on the difference in the neutron index of refraction of the two materials, which is proportional to density and therefore changes with the liquid or gas state of the helium.  These differences can cause systematic effects through target-dependent neutron beam intensity and phase space changes  coupled to residual magnetic fields in the target region \cite{Ba108}. This subclass of neutron trajectories was prevented from reaching the $^{3}$He ionization chamber by collimation of the beam.

Within each target chamber, a set of three $^{6}$LiF-plastic collimators prevented neutrons from reflecting off the chamber walls and reaching the detector as shown in Figure~\ref{collimators}. Collimators were positioned at 1/4, 1/2, and 3/4 of the length of the target chambers and extended into the target chamber 5 mm along the top, bottom, and outer chamber walls, and 2 mm along the chamber septum.  This collimation defined the sub-beam within a target chamber as 26.5 mm wide by 50 mm tall.

The incident neutron beam possesses a broad energy spectrum, which begins at the cold source as a Maxwellian distribution corresponding to a temperature of 40 K.  Because the critical angle of the guides that transport the neutrons to the apparatus increases with the neutron wavelength, any particular choice of collimation suppresses reflected neutrons only below some cutoff wavelength. The geometry and spacing of the collimators sufficed to prevent neutrons with wavelengths less than 2 nm from reflecting off target walls and being accepted by the ionization chamber. The neutron beam intensity for wavelengths above 2 nm in a long ($\sim$60 m) guide like that used at NCNR is typically over three orders of magnitude smaller than the higher energy portion of the beam. This fraction of the beam is too small to make a significant contribution to the systematic uncertainty in the measurement.

The collimators were built from $^{6}$LiF-plastic that was glued to an aluminum backing with Stycast 2850FT epoxy resin and then glued into the target chambers. Additionally, $^{6}$LiF-plastic was glued to the upstream face of each target to define the left and right-side neutron sub-beams with the same collimation dimensions as those inside the target chambers.

\subsection{Pi-Coil}

The pi-coil generated an internal magnetic field that precessed the transverse component of neutron spin about the $\hat{\rm{y}}$-direction.  The amount of spin precession was determined by the strength of the magnetic field and the neutron velocity.  The pi-coil was tuned to rotate the mean wavelength (approximately 0.5 nm) of the neutron beam by $\pi$ rad.

\begin{figure}
    \centering
    \includegraphics[width=\linewidth]{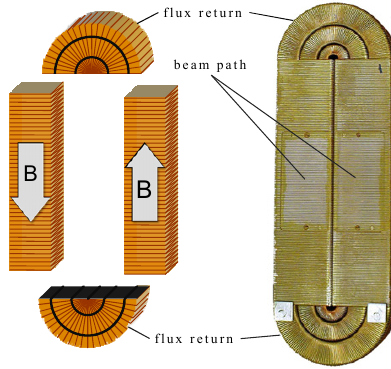}
    \caption{Diagram of the pi-coil.  The coil is wound so that the left and right solenoids produce vertical magnetic fields through the region of the passing neutron beam.  These fields are oriented opposite to each other, and the curved solenoid sections on top and bottom provide flux return for both coils.}
    \label{picoil}
\end{figure}

Designed and constructed by the University of Washington, the pi-coil consisted of a pair of side-by-side, 40 mm square cross-section, 160 mm tall solenoids (see Figure~\ref{picoil}). To minimize magnetic flux leakage, the current in the two rectangular coils flowed in opposite directions with a set of three curved solenoids providing magnetic flux return at the ends.  The leakage field was measured at less than 50 nT at a position of 1 cm from the center of the coil.  Each solenoid in the pi-coil was wound around an aluminum core with three layers of 28 gauge copper magnet wire at a winding density of 10 wires per cm per layer.

The pi-coil was fixed in place between the upstream and downstream target chambers by nylon set screws that were bolted into a surrounding aluminum mount.  The screws pressed onto aluminum relief plates that were glued to the outside of the pi-coil in order to prevent windings from being damaged.  The mount was attached to the upstream and downstream target chamber lids.  A brass pin was fit into the bottom flux return of the pi-coil and coincided with its geometric vertical axis.  The pin rested in a transverse groove in the target support rail, which also secured the target chambers.

The height of the support pin and the positions of the set screws were chosen to center the pi-coil in the beam as defined by the target chamber collimation.  The geometric axis of the pi-coil was positioned to coincide with the vertical axis of the target as defined by the partition between target chambers.  The pi-coil was positioned equidistant between the targets approximately 20 mm from the inside surface of the target chambers.

\subsection{Centrifugal pump and drain system}

Isolating the parity-violating component of the neutron spin rotation through target motion required a method of changing target states that filled and drained diagonal pairs of target chambers without changing the magnetic fields inside the target region.  This was accomplished with a centrifugal pump that was immersed in a 13 L liquid helium bath located in the bottom of the vac-canister and a set of flexible drainpipes that were connected to outlets located at the bottom of each target chamber.

\begin{figure*}
    \centering
    \includegraphics[width=\linewidth]{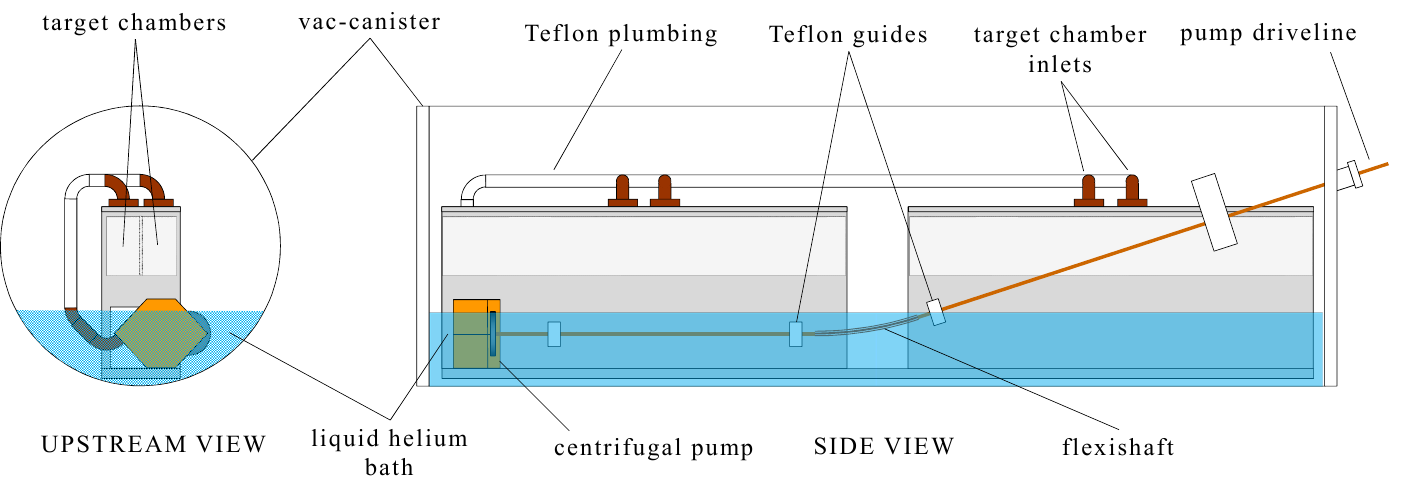}
    \caption{Schematic of the centrifugal pump system.  A stepper motor positioned outside the target region provided torque to the centrifugal pump along a driveshaft in the motion-control feedthrough tube and a driveline in the vac-canister.  The centrifugal pump moved liquid helium from a bath in the bottom of the vac-canister into plumbing that connected to the four target chamber inlets located in the target lids.}
    \label{pump}
\end{figure*}

The centrifugal pump was similar in size and design to positive-displacement pumps that have previously been used to move both normal and superfluid helium \cite{McC73}.  The pump was tested thoroughly in liquid nitrogen at IUCF before operation at liquid helium temperature.  The torque needed to spin the impeller was transferred from a stepper motor located outside of the target region through a carbon fiber driveshaft within the motion-control feedthrough tube and through a driveline inside the vac-canister that was built from brass rod and flexible copper braided-wire rope (Figure~\ref{pump}).

In the first version of this experiment, the centrifugal pump was mounted in a low and horizontal position within the liquid helium bath, such that the impeller spun around a vertical axis.  This arrangement allowed the pump to fill the target chambers in about 30 s, but required a 6:1 gear ratio and a 90$^\circ$ miter gear pair to transfer torque from the horizontal driveshaft to the vertical impeller shaft.  However the pump suffered mechanical failures due to ice impurities within the liquid helium jamming the gears.

Based on the previous performance, the pump was mounted in a vertical position within the liquid helium bath, so that the impeller spun around a horizontal axis.  This allowed the removal of the 90$^\circ$ miter gear pair.  Also, a larger-toothed 4:1 gear ratio was chosen that would be more resistant to seizing due to ice crystals.

The volume throughput of the centrifugal pump was a function of impeller speed and the depth of the liquid helium bath.  The maximum rotation frequency of the stepper motor feedthrough was 300 rev/min.  The centrifugal pump had a 4:1 gear ratio, which set the maximum impeller rotational speed at 120 rev/min.  In practice, operating the stepper motor above 120 rev/min caused the pump and driveline to seize. This rotation speed could fill all four target chambers with liquid helium in 250 s to 300 s, depending on the depth of the bath.

\begin{figure*}
    \centering
    \includegraphics[width=\linewidth]{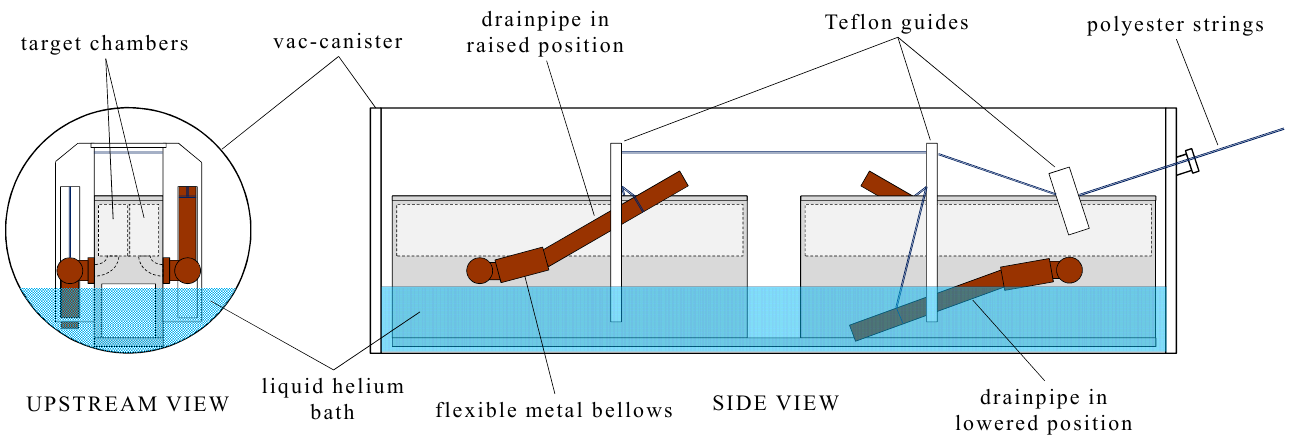}
    \caption{Schematic diagram of the drain system.  Actuators located outside the target region move strings connected to the ends of drainpipes that are attached to the outlets of the target chambers via flexible metal bellows.  The raised or lowered position for each drain was chosen to ensure that the tip of the drainpipe was above the full liquid level or below the empty level (respectively) for a target chamber.}
    \label{drains}
\end{figure*}

Target chambers were emptied of liquid helium by drainpipes that could be raised or lowered by braided polyester strings, which were routed through the target and motion-control feedthrough tube through Teflon and carbon fiber sleeves to a pair of pneumatic linear actuators located outside of the target region.  Drainpipes from diagonal pairs of target chambers (e.g. upstream-right and downstream-left) were operated together from a single actuator and could empty a target chamber of liquid helium in 50 s when lowered (Figure~\ref{drains}).

Inside each target chamber, square grooves were machined into the floor and lid to ensure that liquid helium or helium gas did not collect along the bottom or top (respectively) of the chambers between collimators.  This situation could have created a helium liquid/gas interface, which might have reflected some neutrons that would otherwise have been stopped by the collimation and could have been counted in the $^{3}$He ionization chamber.

\subsection{Pyroelectric Ice Getters}

The centrifugal pump had gears and other moving parts that could in principle be jammed by solid impurities mixed within the liquid helium.  Possible impurities include ice crystals from water or liquid air accidentally introduced into the target system during a liquid helium transfer or from the ice slurry found in the bottom of a typical liquid helium research dewar. These impurities caused the gears of the pump to seize in the first version of the experiment.  We decided to protect the centrifugal pump with a pyroelectric object.

Cesium nitrate is a ferroelectric that spontaneously polarizes at cryogenic temperatures \cite{Bro89} and remains polarized at constant temperature in a cryogenic environment.  Cesium nitrate powder was mixed with urethane resin and cast into discs 75 mm in diameter by 5 mm thick.  The discs were bolted onto the lower section of each target, so they would be immersed in the liquid helium bath.

Although we made no attempt to measure the impurities that may have been present in the liquid helium inside the target, based on our experience it is quite unrealistic to assume that they were absent. In our judgement, the most likely explanation for the fact that the same centrifugal pump used in the previous experiment did not seize over several months of nearly continuous operation is the presence of these getters along with gear modifications described in Section 2.5.

\section{Motion Control System}

In order to change target states, liquid helium needed to be non-magnetically moved between target chambers.  The liquid helium target system employed a pump and drain system that was operated outside of the target region by the motion-control system.

The motion-control system consisted of a stepper motor and driveshaft, which turned the centrifugal pump, and a pair of pneumatic linear actuators that were attached to the drains by strings.  The stepper motor and actuators were connected to a vacuum chamber called the motion control box (MCB), which shared the same helium environment as the target system inside of the vac-canister.  A motion-control feedthrough tube guided the driveshaft and actuator control strings from the MCB into the vac-canister through penetrations in the magnetic shielding and cryostat.

\begin{figure*}
    \centering
    \includegraphics[width=\linewidth]{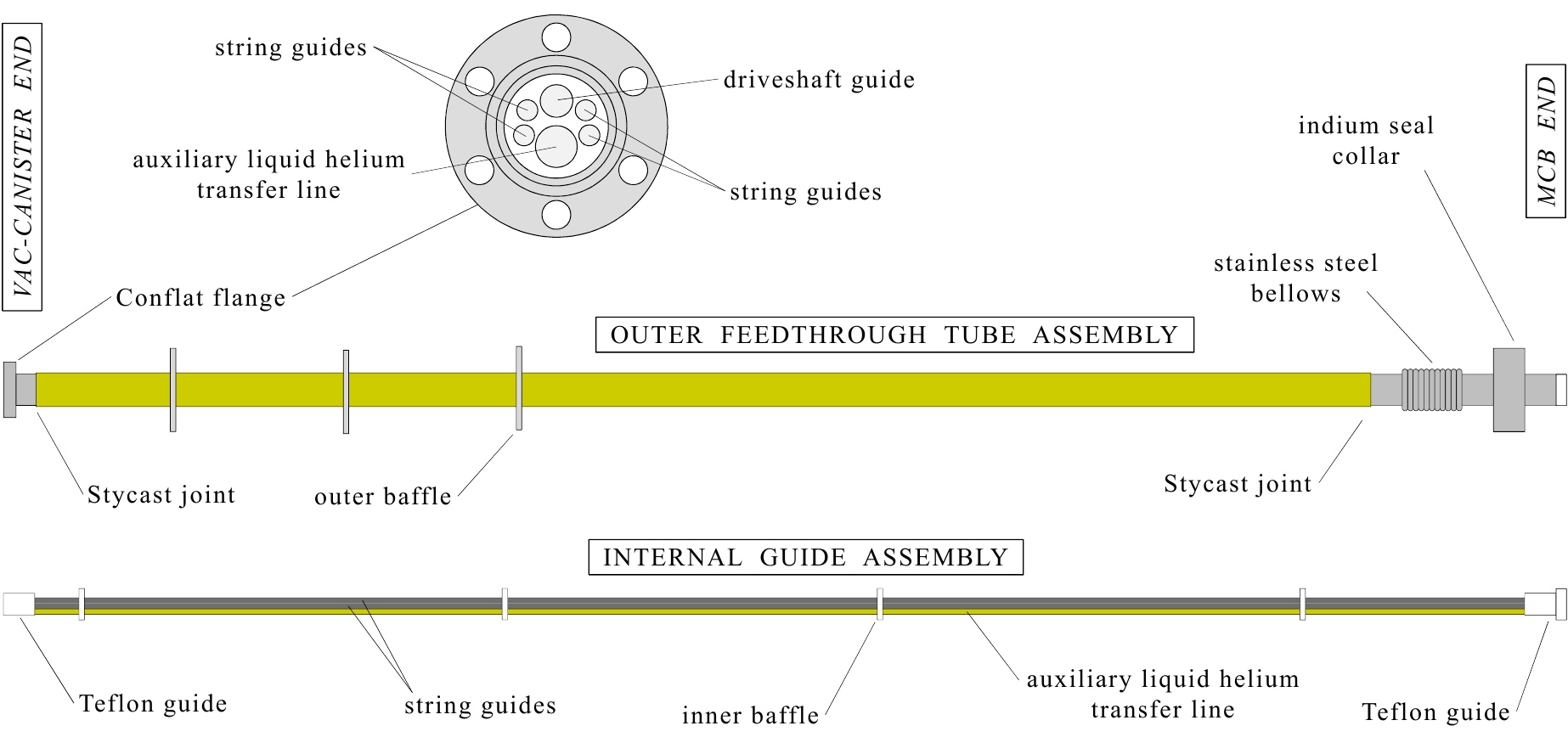}
    \caption{Schematic diagram of the motion control feedthrough tube.  Four carbon fiber tubes that sheathed the drainpipe strings were glued into a set of thermal radiation baffles.  A hole in each baffle guided the pump driveshaft.  The bundle fit within a laminate tube and was capped by Teflon guides on each end.  The laminate tube was externally coated with a layer of epoxy resin and reflective aluminum tape.  A set of baffles were glued on the outside of the laminate tube to suppress radiative heat loads on the vac-canister. An aluminum CF flange and a stainless steel bellows assembly were glued into the ends of the feedthrough tube.}
    \label{tubes}
\end{figure*}

The motion-control feedthrough tube was constructed from thin-walled G10-grade glass epoxy laminate tube, which was coated with epoxy resin and surrounded by a layer of reflective aluminum tape.  The epoxy resin provided additional mechanical stiffness to the laminate tube and suppressed helium diffusion and light transmission through the wall of the laminate.  The aluminum tape provided a highly reflective surface that suppressed radiative heat transfer by thermal radiation from the aluminum vacuum jacket that surrounded the motion-control feedthrough tube.  An aluminum CF flange was glued into the cold end of the tube, and a 316 stainless steel bellows assembly was glued into the room temperature  end, which was located outside of target region (Figure~\ref{tubes}).

A guide assembly was housed within the length of the motion-control feedthrough tube and was built from four small diameter carbon fiber tubes that were glued into a set of baffles.  The carbon fiber tubes sheathed the braided polyester control strings that connected the target chamber drains to actuators within the MCB.  The baffles provided mechanical support to the carbon fiber tubes and allowed the guide assembly to slide within the motion-control feedthrough tube under differential thermal contraction.  The baffles also prevented light and thermal radiation from shining onto the interior of the vac-canister and segmented the helium gas column within the motion-control feedthrough tube to suppress potential  heat loads due to gas convection.  A drilled hole in each baffle supported and aligned the carbon fiber driveshaft, which connected the centrifugal pump driveline in the vac-canister to the stepper motor located in the MCB.  Teflon caps were inserted into each end of the motion-control feedthrough tube and provided a smooth bearing surface for the strings and driveshaft.

The driveshaft coupled to the stepper motor inside the MCB via a stainless steel double universal joint.  The control strings attached onto the ends of actuators through brass tension springs.  The lengths of the strings were chosen to ensure that each drain would travel through its entire range of movement, and the springs provided tension relief for the strings.  The springs were chosen so that they would mechanically fail before a string broke in case of a stuck string. The MCB possessed a large access port with a Buna-N o-ring seal and an acrylic window for both visual inspection and (if needed) mechanical repair of the springs, strings and actuators.

The aluminum vacuum jacket that surrounded the motion-control feedthrough tube coupled into the MCB through a fluorosilicone compression o-ring seal.  This type of material remains plastic through a wider temperature range than typical silicone or fluorocarbon o-rings \cite{Par07}, which was important since the MCB became cold during target operation.  During liquid helium transfers or other periods of rapid cryogenic liquid boil-off in the target, heater tape was used to guard against the development of vacuum leaks due to o-ring embrittlement caused by excessive cold.

\section{Cryogenics}

\subsection{Cryostat}

A horizontal bore, nonmagnetic cryostat was originally built by Oxford Instruments for a previous experiment to measure the parity-violating neutron spin rotation in liquid helium.  All of the vacuum joints that were originally glued had failed, and so new joints were redesigned and replaced with either soldered or indium-sealed joints as applicable.

The cryostat consisted of two coaxial annular aluminum vessels housed within an aluminum cylindrical main vacuum vessel. The outer 77-K vessel could hold 50 L of liquid nitrogen, and the inner 4-K vessel could hold 30 L of liquid helium. The measured hold time of the cryostat during nominal data runs was 50 h for liquid nitrogen and 30 h for liquid helium, which allowed a convenient daily refill schedule.

The cylindrical interior surface of the 4-K vessel formed the cold bore, which was 305 mm in diameter by 1000 mm long.  A cryogenic magnetic shield built from Amuneal Cryoperm 10 lined the cold bore.  Cryoperm 10 was chosen because its permeability at cryogenic temperatures is comparable to that of normal mu-metal at room temperature.

All materials of the cryostat that were located inside the room-temperature magnetic shielding were nonmagnetic. The two annular vessels were independently suspended within the main vacuum vessel by adjustable G-10 straps and braces for thermal isolation and adjustability.  The cryostat was supported within the magnetic shielding by a set of four aluminum posts, which passed through small openings in the shielding.  The posts fit inside machined inserts and supported the cryostat from its ends.

\subsection{Liquid Helium Transfer Tube}

The liquid helium target was subject to a heat load that boiled away the liquid helium in the vac-canister, thereby decreasing the depth of the liquid helium bath over time.  A minimum bath depth was required for changing target states, so the vac-canister was periodically refilled from an external dewar using a liquid helium transfer tube that entered  the vac-canister through an external valve and compression o-ring fitting assembly located outside of the target region and magnetic shielding.

The liquid helium transfer tube was of similar design as the motion-control feedthrough tube (see section 3), except that there was no internal guide assembly.  An aluminum vacuum jacket surrounded the liquid helium transfer tube.

\subsection{Heat Load}

The boil-off rate of liquid helium in the vac-canister determined the upper bound on the length of a data run.  Suppressing this heat load allowed longer data runs and therefore greater statistical accuracy for the spin rotation measurement. Considerable effort was therefore devoted to the reduction of possible heat leaks in the system. The challenge was to achieve low heat leaks in a liquid helium target system which necessarily possesses direct mechanical linkages between room temperature and the inside regions of the vac-canister.

Electrical instrumentation in the target was turned off when not required in order to reduce the heat load generated by current-carrying wires and sensors.  This included turning off the pi-coil during target changes.

All cryogenic surfaces were either polished or layered with reflective aluminum tape or superinsulation to suppress radiative heat transfer.  Aluminum heat shields were bolted onto the upstream and downstream annular faces of the cryostat liquid helium vessel, which provided nearly 4$\pi$ coverage of the vac-canister surface with a 4 K surface.  A second set of heat shields anchored to the cryostat liquid nitrogen vessel provided a 77 K surface exterior to the 4 K surface.  Beam windows on these heat shields were covered with aluminum foil, and the inner face of each shield was covered with cryogenic super-insulation.

Both the motion-control feedthrough tube and the liquid helium transfer tube penetrated the 4-K and 77-K heat shields and were sheathed by a pair of aluminum vacuum jackets outside of the target region.  The outer surface of these tubes were wrapped with super-insulation, and each of them employed a set of exterior baffles that suppress 300 K thermal radiation from being incident on the vac-canister through openings in the heat shields.

All components anchored to a surface warmer than 4 K were constructed using materials that possessed low thermal conductivity and were sufficiently long to suppress conductive heat transfer.  The wiring harness for the target instrumentation was thermally lagged to the heat shields.  Similarly, the motion-control feedthrough tube and the liquid helium transfer tube were thermally lagged to the downstream 4-K and 77-K heat shields with braided copper straps.

The design estimate for the total heat load on the liquid helium target was 100 mW to 150 mW.  However, this estimate was too small to explain the observed bath boil-off rate of about 10 L of liquid helium every 6 h to 8 h.  A numerical simulation was conducted of the thermal transport along the motion control feedthrough tube and the liquid helium transfer tube, which revealed that the layer of reflective aluminum tape wrapped around the outer surface of each tube conducted 140 mW to 225 mW from the MCB and liquid helium valve transfer port at room-temperature to the vac-canister at 4 K.  Including the effects of the radiative heat load along the length of each tube from the surrounding room-temperature vacuum jackets -- which was what the aluminum tape was introduced to suppress -- added another 20 mW.

\begin{figure}
    \centering
    \includegraphics[width=\linewidth]{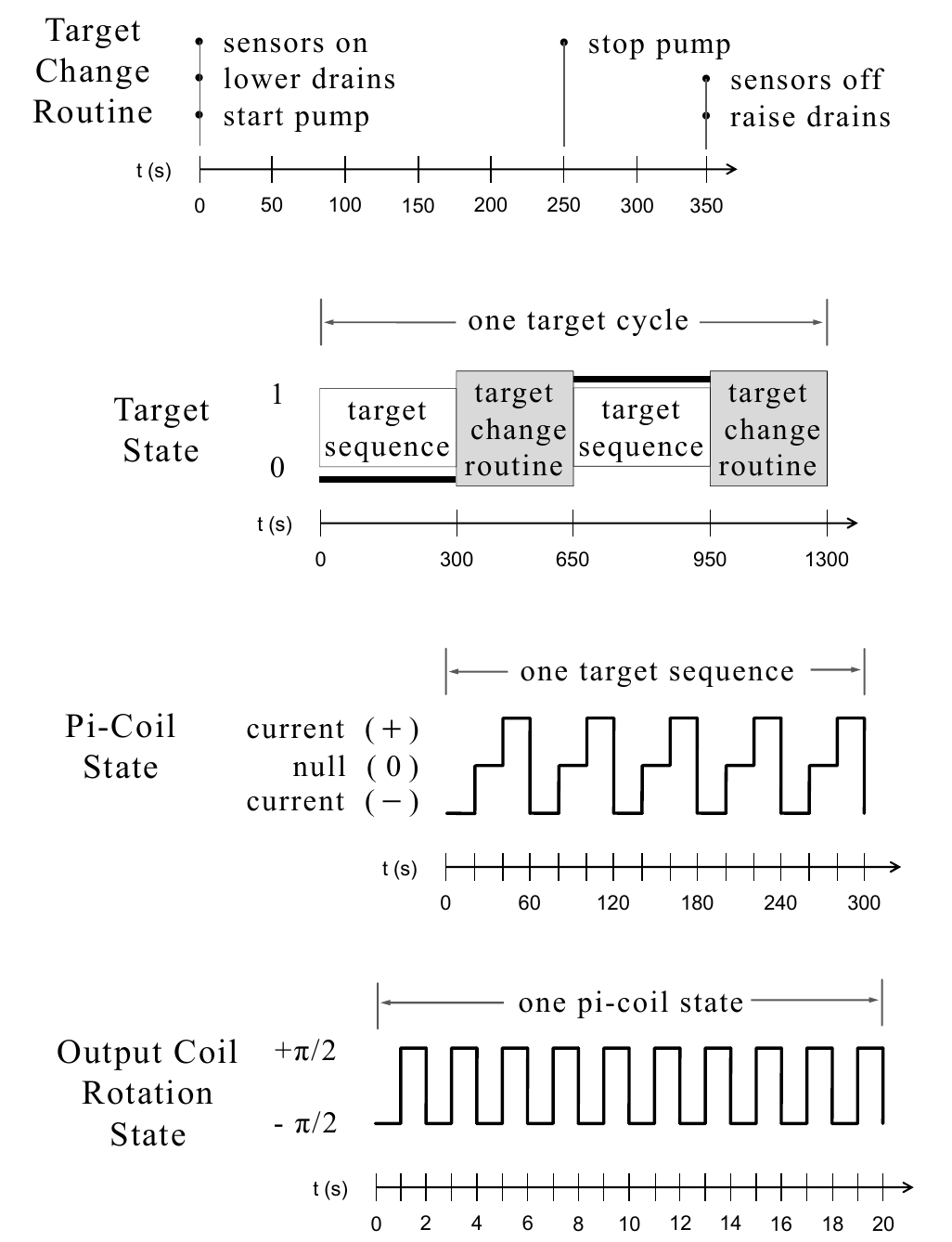}
    \caption{Diagram of the target control sequence.  Once data-taking for one target state has completed, the \emph{target-change routine} initiates, which moves liquid helium between target chambers and configures the target in the next target state.  Data-taking continues by running the target through a \emph{target sequence}, which involves stepping the pi-coil state through five series of current configurations (both directions of current within the pi-coil and a null current setting).  During each pi-coil state, the helicity of the output coil field modulates 20 times at a rate of 1 Hz.  A complete \emph{target cycle} consists of two consecutive \emph{target sequences} (one for each target state) with the two intervening \emph{target-change routines}.}
    \label{routine}
\end{figure}

In addition, the simulation indicated that the heat load along the tubes due to the 77-K thermal lagging straps was substantial, because the copper braids mechanically coupled onto each tube about 20 cm from the face of the vac-canister.  The straps were anchored to the 77-K heat shield, which could be tens of degrees warmer than the cryostat 77-K vessel to which it was coupled.  The 4-K thermal lagging straps, which connected onto the tubes about 10 cm from the vac-canister, partially offset the heat load due to the 77-K straps.  But, the 4-K straps were anchored to the 4-K shield, which was several degrees warmer than the cryostat 4-K vessel to which it was attached.

The simulation suggested that the motion-control feedthrough tube with its internal guide assembly, the pump driveshaft and control strings, and the LHe transfer tube allowed 450 mW to 750 mW of heat into the vac-canister.  This heat load is close to the amount of heat needed to account for the helium boil-off observed during data runs.  Any future operation of the apparatus will include design changes that would decrease this heat load to an amount closer to the original design.

\section{Experimental Control and Data Acquisition}

\subsection{Instrumentation}

The liquid helium levels for each target chamber and the bath in the vac-canister were individually monitored with resistive-wire liquid level sensors.  The temperatures on the upstream and downstream target, as well as the cryostat 4-K and 77-K vessels were monitored with silicon diode temperature sensors.  The longitudinal magnetic field within the target region was measured by a fluxgate magnetometer with a stated sensitivity of 0.5 nT.  The magnetometer was multiplexed to four single-axis low-temperature probes that were mounted above the targets.

\subsection{Experimental Control}

The control sequence for the liquid helium target system, the data acquisition, and data storage were managed by the Neutron Spin-Rotation Acquisition and Control (NSAC) program.  The NSAC collected neutron beam count rates from the ionization chamber while it switched through the combinations of the two output coil rotation states ($+\pi/2$, $-\pi/2$) with the three pi-coil current states (current[-], off[0], current[+]) in a predetermined sequence.  This sequence was iterated several times to build up a \emph{target sequence} (see Figure~\ref{routine}).

\begin{figure*}
    \centering
    \includegraphics[width=\linewidth]{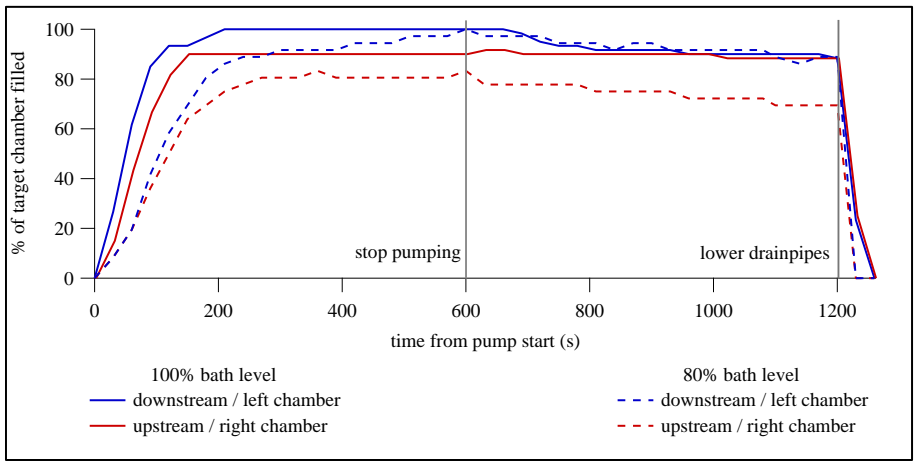}
    \caption{Plot of the liquid helium level in a target chamber as measured by a resistive wire liquid helium meter during fill and drain testing for various depths of the liquid helium bath in the vac-canister.  The deepest bath depth produced the fastest fill curve, while decreases in the bath corresponded to slower fill times.}
    \label{levels}
\end{figure*}

After a target sequence was completed, the NSAC stopped collecting beam rate data and initiated a \emph{target-change routine}.  The temperature sensors and one of the fluxgate magnetometer sensors were energized and allowed to warm up to operational temperature.  At the same time, a motion control actuator lowered the pair of drains of the target chambers that were full of liquid helium during the preceding target sequence and allowed them to empty their liquid helium into the bath.  Then, the stepper motor began spinning the centrifugal pump, which began filling all four target chambers with liquid helium from the bath.

The NSAC turned the stepper motor off after a given length of time, which allowed the pump to completely fill the pair of target chambers with the raised drains with liquid helium.  Calibration testing of the centrifugal pump using varying depths of liquid helium in the bath indicated that stepper motor run times of 300 s was sufficient for target switch-overs that occurred in the first half of an eight-hour data run and 350 s for the later half.  The difference in pump times was related to the decreasing bath depth over the duration of a run from liquid helium boil-off.

Once the centrifugal pump had stopped, data from the temperature and fluxgate sensors were recorded by the NSAC.  The liquid helium was allowed to completely empty from the target chambers with lowered drains for 50 s, a time which was determined during calibration measurements.  This drain time also allowed any bubbles within a full target chamber to settle out so that the liquid helium density was homogeneous during neutron data sequences.  The drain time also allowed the temperature in the targets to equilibrate and any turbulence in the target chambers to subside.

After the empty target chambers drained, the actuators raised the drains and all instrumentation in the target region was powered off.  With the possible exception of the angular orientation of the centrifugal pump vanes and other rotating elements along the driveline, all of the locations of the moveable mechanical systems inside the target chamber are in the same location for all data runs after the liquid motion sequence.  The target was now in the new \emph{target state}, and data taking could resume for the next \emph{target sequence}.  Two consecutive target sequences -- one for each target state -- constituted a \emph{target cycle}.  Upon completion of each target cycle, the NSAC calculated various count rate asymmetries for the target, total spin rotation angles, and parity-violating spin rotation angles $\phi_{\rm{PV}}$ for real time analysis.

Because the target-change routine was constrained by hardware performance, the target cycle duration was determined by the number of modulations of the output coil rotation states for each pi-coil state, the frequency of the modulation, and the number of pi-coil sequences within a target sequence.  The frequency choice for the output coil was discussed in Section 1.3, and the number of modulations determined the statistical precision of the asymmetry measurement for a single pi-coil state.

The concern for drifting magnetic fields in the target region placed a bound on the number of pi-coil sequences in a target sequence, because the asymmetry measurements for a given pi-coil state could include background rotation signals due to different background magnetic fields, which would reduce the precision of the measurement.  Similarly, because the parity-violating spin rotation angle $\phi_{PV}$ is calculated from the asymmetries from the same pi-coil states for each target state in consecutive target cycles, a drifting magnetic field could reduce the statistical precision of the measurement or introduce a systematic error.

However, the duration of the target-change routine introduced a significant dead time during data-taking.  The length of the target sequence was chosen to limit dead time and increase statistics while suppressing possible systematic errors and increasing the precision of the measurement.  Therefore, each target sequence consisted of five iterations of each pi-coil current state, each of which contained 20 modulations of the output coil rotation state with a frequency of 1 Hz.  A complete target cycle of two target sequences and two target-change routines lasted 1300 s.

\begin{figure*}
    \centering
    \includegraphics[width=\linewidth]{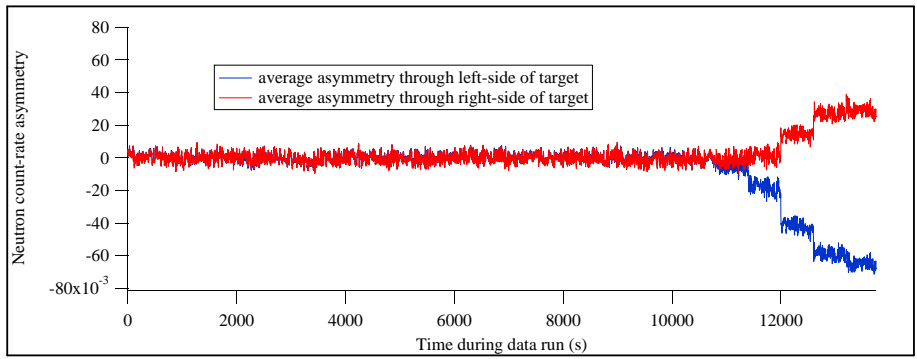}
    \caption{Plot of the neutron count-rate asymmetries for the left and right sides of the target during a run in cycle 2.  The pattern at the end of the run indicates partial filling of target chambers due to decreasing bath depth in the vac-canister.  The pattern's step-size of 300 s corresponds to the modulation of the target states.}
    \label{asymmetry}
\end{figure*}

\section{Liquid Helium Target System Performance}

The liquid helium target system was assembled and tested at IUCF.  Basic testing of the target was conducted to ensure that instrumentation and motion control components functioned correctly at low temperatures.  The apparatus was then shipped to the NCNR, and after initial beamline and polarimeter characterization studies were complete, the liquid helium target system was installed and commissioned on the NG-6 beamline.

During the commissioning, neutron rates were measured at various points along the apparatus while the liquid helium target system was run in an operational target state.  In addition, rates were measured for an empty target configuration, where the target chambers were drained, but liquid helium was present in the bath.  The neutron rates for the target are summarized in Table 1.

\begin{table}
  \begin{center}
    \begin{tabular}{l c}
        & Peak fluence (neutrons cm$^{-2}$ s$^{-1}$) \\
        \hline
        \\
        After supermirror polarizer & $3.1\times10^{8}$ \\
        Before polarization analyzer (target empty) & $2.8\times10^{7}$ \\
        Before polarization analyzer (target full) & $2.3\times10^{7}$ \\
    \end{tabular}
    \label{fluence}
    \caption{Typical neutron rates for the liquid helium target.}
  \end{center}
\end{table}

The liquid helium target operated more or less continuously during the experiment for about six months in early 2008, executing 5406 target motion sequences with liquid helium.  The target was warmed once to room temperature during a reactor shutdown to fix a small intermittent leak at the low temperature end of the G-10 tube used to repeatedly transfer liquid helium into the vac-canister.  Except for this warm-up, the target was always held at a temperature no greater than 77 K to minimize the development of internal stresses on seals from differential thermal contraction.

The performance of the target fill and drain system in moving liquid between target chambers was tested prior to the experiment.  Figure~\ref{levels} shows the liquid helium levels as measured by the liquid level meters in the course of a fill and drain sequence similar to that used in the experiment.  The data demonstrated that the centrifugal pump worked and that the drain pipes operated as designed. It should be noted that the testing revealed that the downstream target chambers filled about 25\% quicker than the upstream chambers.  This was likely due to the downstream location of the centrifugal pump and the additional length of plumbing needed to transport liquid helium to the upstream chambers.

During the experiment, the liquid helium levels within the target chambers could be monitored indirectly using the neutron beam transmission.  The count-rate asymmetry used to calculate the neutron spin-rotation angle was separated into left and right-side count rate asymmetries.  Any failure of the target to either fill or drain properly was indicated in the neutron data as a large deviation of these left--right count rate asymmetries from zero.

Figure~\ref{asymmetry} shows a plot of count rate asymmetries during a typical data run.  The neutron data indicate that initially, the liquid helium bath was overfilled and that the drains didn't fully empty all of the chambers between target sequences.  After several target sequences, normal fill and drain performance was indicated.  Towards the end of the run, the neutron data indicate that the target chambers were not filling or draining properly, which coincided with the depth of the liquid helium bath decreasing below nominal operating levels due to boil-off.

The target could also be operated without liquid helium present in the target chambers or even the target chambers and vac-canister.  This was done before and after the liquid helium phase of data taking to place constraints on possible systematic effects.

\begin{figure*}
    \centering
    \includegraphics[width=\linewidth]{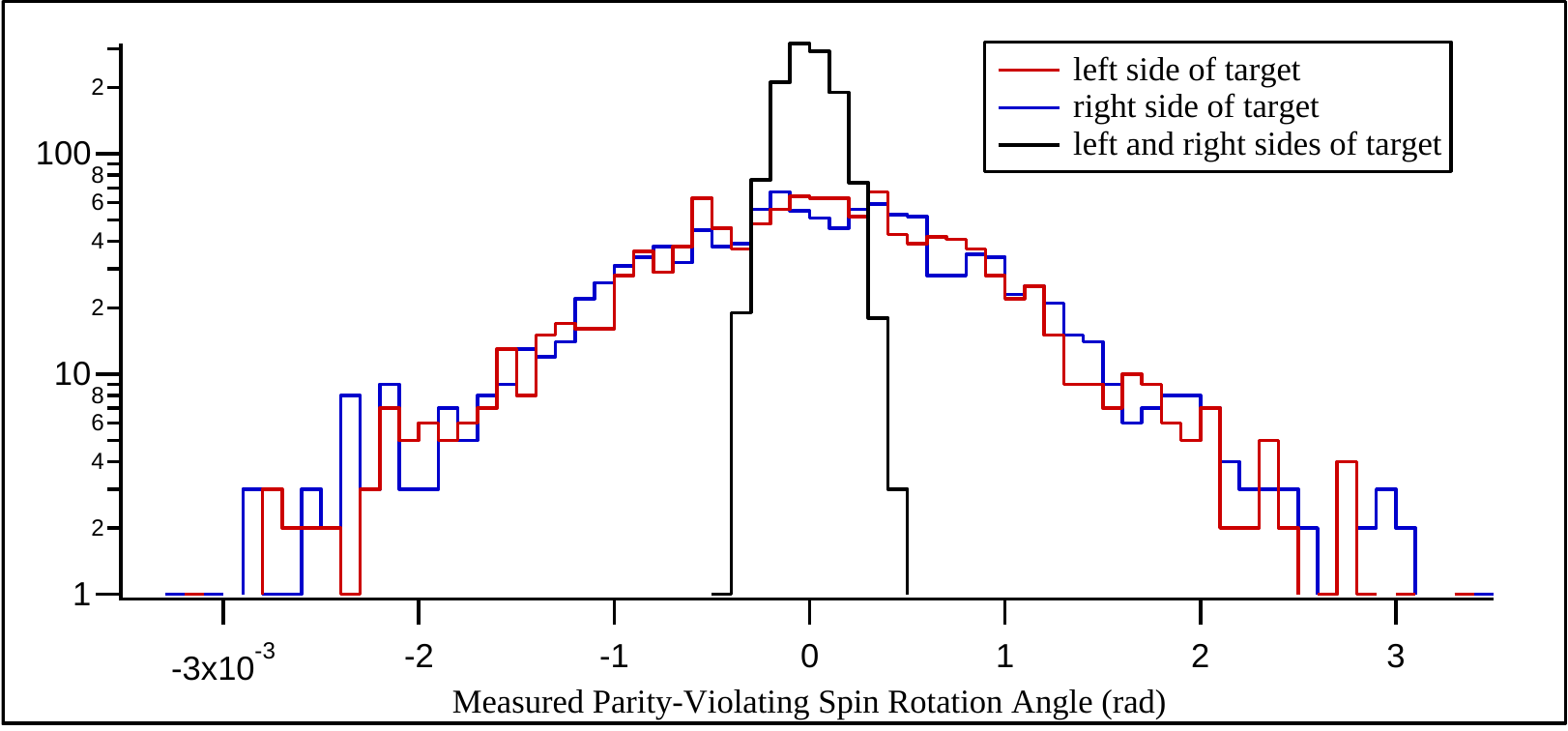}
    \caption{Histogram of the measured parity-violating spin rotation angles with and without common-mode noise suppression.}
    \label{noise}
\end{figure*}

The left-right segmentation design of the target allowed a suppression of common-mode noise due to the intensity noise from the reactor described in Section 1.3.  As shown in Figure~\ref{noise}, this suppression increased the measurement precision by roughly an order of magnitude, with statistical uncertainties approaching $\sqrt{N}$.

\begin{figure*}
    \centering
    \includegraphics[width=\linewidth]{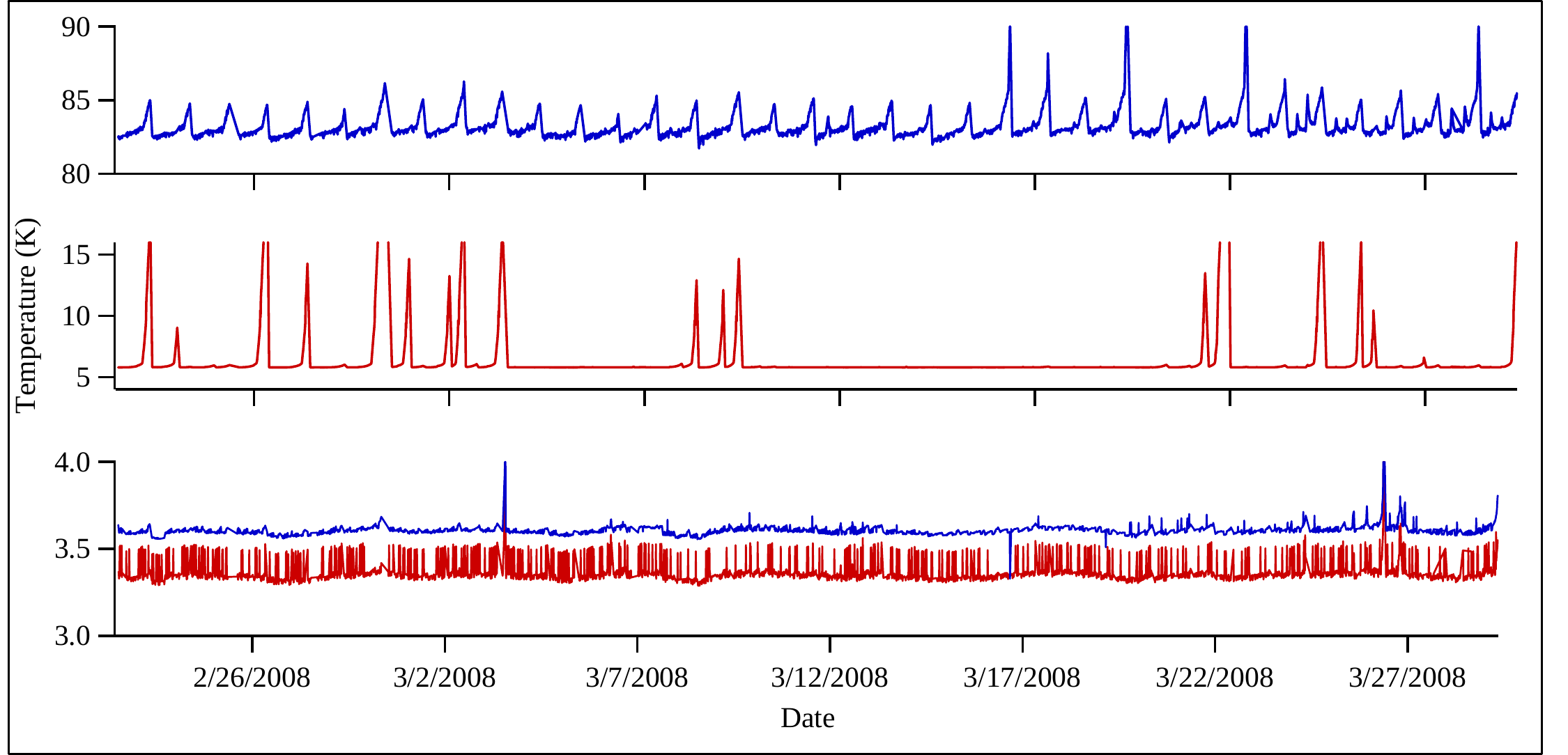}
    \caption{Plot of measured temperatures during cycle 2 of the liquid nitrogen jacket, the liquid helium jacket, and the downstream and upstream target chambers (shown in order top to bottom).  The liquid nitrogen jacket displays daily warming and cooling that coincide with its daily fill schedule.  The liquid helium jacket shows good temperature stability, with the noted exception of temperature spikes during some of the daily fills.  The target temperatures similarly display temperature spikes during liquid helium fills, but at a three time per day frequency that coincided with the scheduled 8-hour runs; otherwise, the target temperatures displayed good stability during data runs.}
    \label{temperature}
\end{figure*}

The temperature of the target was measured periodically throughout the run during target motion.  Figure~\ref{temperature} shows the temperatures as measured by thermometers located at various locations on the aluminum target chamber.  No evidence was observed for excursions of the target temperature away from that expected for a liquid helium bath whose pressure is close to atmosphere.

\section{Conclusion}

The Neutron Spin Rotation experiment acquired data from January 2008 to June 2008.  During the experiment, the liquid helium target system met or exceeded most design goals. In particular, the motion control and centrifugal pump systems performed reliably in the cryogenic environment.  The drain system worked especially well, with drain times as expected or better.

Further experiments to perform precision measurements of neutron spin rotation in liquid helium are possible using the same polarimeter and target system design.  Removal of the reflective aluminum tape on the motion control feedthrough tube and the liquid helium transfer tube and a redesign of the thermal anchoring for these tubes should greatly reduce the liquid helium consumption rate of the target.

Another improvement in target performance could come from increasing the centrifugal pump speed, thereby reducing the time needed to fill target chambers with liquid helium.  This could be accomplished by returning the pump to its original low and horizontal position within the liquid helium bath, choosing appropriate large-toothed gears, continued use of the ice-getters, and development of a ``cage'' that would surround the gear system from any remaining ice crystals.  This pump modification could decrease the fill times to the 50 s to 100 s range, which could decrease the time needed for a target change routine by a factor of three.  This reduction of dead time would increase the available neutron count rate statistics accordingly.

In addition to measurements done with liquid helium, other targets are possible, including liquid parahydrogen \cite{Avi84A}\cite{Mar05}\cite{Sch04} and liquid orthodeuterium \cite{Avi84B}\cite{Sch08}.  These targets would require modifications of the basic liquid helium target system for the warmer operational temperatures (20 K) and the inclusion of hydrogen safety systems.

\section{Acknowledgements}

The authors acknowledge the support of the National Institute of Standards and Technology, U.S. Department of Commerce, in providing the neutron research facilities used in this work, and the support of the U.S. Department of Energy, Office of Nuclear Science and the National Science Foundation.

We thank the University of Washington for funding and use of hardware from the previous Neutron Spin Rotation experiment.  We thank the Indiana University Physics Department machine shop and the University of Washington machine shop for their work on fabrication of the liquid helium target system components, and the Indiana University Cyclotron Facility for infrastructure used in the extensive target testing conducted before the experiment.  We thank David Haase for his work in refurbishing the cryostat and room-temperature magnetic shielding.  Christopher Bass acknowledges the support of the Ronald E. McNair Graduate Fellowship Program and the National Research Council Postdoctoral Associates Program.  This work is supported in part by NSF PHY-0457219, NSF PHY-0758018, NSF PHY-9804038, and DOE DE-FG02-87ER41020.

\end{document}